\documentclass[aps,prx,reprint,superscriptaddress]{revtex4-2}

\usepackage{amsmath}
\usepackage{amssymb}
\usepackage{amsthm}
\usepackage{braket}
\usepackage{physics}
\usepackage{siunitx}
\usepackage{graphicx}

\newcommand{\bodyfigurelabel}[1]{\label{#1}}
\newcommand{\edfigurelabel}[1]{\label{#1}}
\usepackage{hyperref}

\begin{document}

\title{Scalable Optical Quantum State Synthesizer with Dual-Mode Resonator Memory}
\author{Fumiya Hanamura}
\affiliation{Department of Applied Physics, School of Engineering, The University of Tokyo, 7-3-1 Hongo, Bunkyo-ku, Tokyo 113-8656, Japan}

\author{Kan Takase}
\affiliation{Department of Applied Physics, School of Engineering, The University of Tokyo, 7-3-1 Hongo, Bunkyo-ku, Tokyo 113-8656, Japan}
\affiliation{Optical Quantum Computing Research Team, RIKEN Center for Quantum Computing, 2-1 Hirosawa, Wako, Saitama 351-0198, Japan}
\affiliation{OptQC Corporation, 1-16-15 Minami-Ikebukuro, Toshima, Tokyo, Japan}

\author{Kazuki Hirota}
\affiliation{Department of Applied Physics, School of Engineering, The University of Tokyo, 7-3-1 Hongo, Bunkyo-ku, Tokyo 113-8656, Japan}

\author{Rajveer Nehra}
\affiliation{Department of Applied Physics, School of Engineering, The University of Tokyo, 7-3-1 Hongo, Bunkyo-ku, Tokyo 113-8656, Japan}
\affiliation{Department of Electrical and Computer Engineering, University of Massachusetts-Amherst, Amherst, Massachusetts 01003, USA}
\affiliation{Department of Physics, University of Massachusetts-Amherst, Amherst, Massachusetts 01003, USA}

\author{Florian Lang}
\affiliation{Department of Information Technology and Electrical Engineering, ETH Zurich, 8092 Zurich, Switzerland}

\author{Shigehito Miki}
\affiliation{Advanced ICT Research Institute, National Institute of Information and Communications Technology, 588-2 Iwaoka, Nishi-ku, Kobe, Hyogo 651-2492, Japan}

\author{Hirotaka Terai}
\affiliation{Advanced ICT Research Institute, National Institute of Information and Communications Technology, 588-2 Iwaoka, Nishi-ku, Kobe, Hyogo 651-2492, Japan}

\author{Masahiro Yabuno}
\affiliation{Advanced ICT Research Institute, National Institute of Information and Communications Technology, 588-2 Iwaoka, Nishi-ku, Kobe, Hyogo 651-2492, Japan}

\author{Takahiro Kashiwazaki}
\affiliation{NTT Device Technology Labs, NTT Corporation, 3-1, Morinosato Wakamiya, Atsugi, Kanagawa 243-0198, Japan}

\author{Asuka Inoue}
\affiliation{NTT Device Technology Labs, NTT Corporation, 3-1, Morinosato Wakamiya, Atsugi, Kanagawa 243-0198, Japan}

\author{Takeshi Umeki}
\affiliation{NTT Device Technology Labs, NTT Corporation, 3-1, Morinosato Wakamiya, Atsugi, Kanagawa 243-0198, Japan}

\author{Warit Asavanant}
\affiliation{Department of Applied Physics, School of Engineering, The University of Tokyo, 7-3-1 Hongo, Bunkyo-ku, Tokyo 113-8656, Japan}
\affiliation{Optical Quantum Computing Research Team, RIKEN Center for Quantum Computing, 2-1 Hirosawa, Wako, Saitama 351-0198, Japan}
\affiliation{OptQC Corporation, 1-16-15 Minami-Ikebukuro, Toshima, Tokyo, Japan}

\author{Mamoru Endo}
\affiliation{Department of Applied Physics, School of Engineering, The University of Tokyo, 7-3-1 Hongo, Bunkyo-ku, Tokyo 113-8656, Japan}
\affiliation{Optical Quantum Computing Research Team, RIKEN Center for Quantum Computing, 2-1 Hirosawa, Wako, Saitama 351-0198, Japan}

\author{Jun-ichi Yoshikawa}
\affiliation{Optical Quantum Computing Research Team, RIKEN Center for Quantum Computing, 2-1 Hirosawa, Wako, Saitama 351-0198, Japan}

\author{Akira Furusawa}
\affiliation{Department of Applied Physics, School of Engineering, The University of Tokyo, 7-3-1 Hongo, Bunkyo-ku, Tokyo 113-8656, Japan}
\affiliation{Optical Quantum Computing Research Team, RIKEN Center for Quantum Computing, 2-1 Hirosawa, Wako, Saitama 351-0198, Japan}
\affiliation{OptQC Corporation, 1-16-15 Minami-Ikebukuro, Toshima, Tokyo, Japan}

\begin{abstract}
Optical quantum computing is a promising approach for achieving large-scale quantum computation. While Gaussian operations have been successfully scaled, the inherently weak nonlinearity in optics makes generating highly non-Gaussian states a critical challenge for universality and fault tolerance. Here, we propose and experimentally demonstrate a scalable method to generate optical non-Gaussian states with a resonator-based quantum memory that supports continuous-time storage and retrieval, in contrast to conventional loop-based memories. We introduce a dual-mode operation of the memory, enabling both storage and entangling functionalities within a single device. By employing a time-domain-multiplexed approach, we successfully demonstrate both cat and Gottesman–Kitaev–Preskill (GKP) breeding protocols in a scalable fashion, marking a key step toward quantum error correction.

Our experiment also marks the first full demonstration of an optical resonator memory performing writing, storage, and readout operations. We validate the memory by storing squeezed single-photon states with up to \SI{93}{\%} total efficiency, and measure an energy relaxation time $T_1 = \SI{2.3}{\micro s}$ and dephasing time $T_\phi = \SI{0.96}{\micro s}$.

These results establish a scalable pathway to generating complex non-Gaussian states required for fault-tolerant optical quantum computing. Beyond computation, our techniques provide new tools for enhancing quantum communication, sensing, and metrology.

\end{abstract}

\maketitle
\section{Introduction}
\begin{figure*}[htbp]
    \centering
    \includegraphics[scale=1]{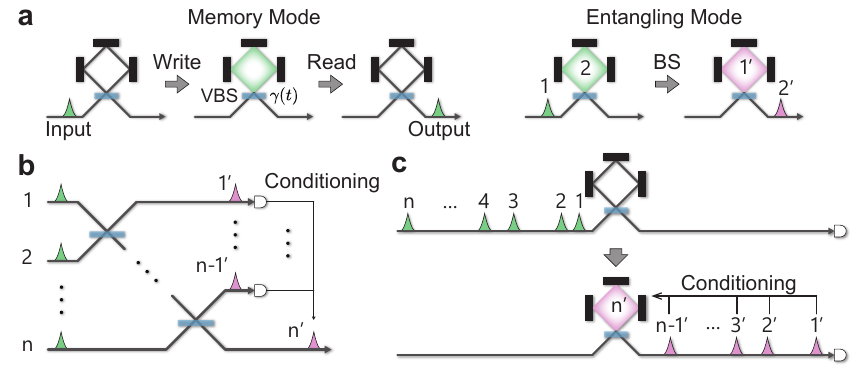}
    \caption{\bodyfigurelabel{fig:dual_mode_breeding} Dual-mode resonator quantum memory and TDM breeding protocol. 
\textbf{a} Dual-mode operation of the resonator quantum memory, modeled as an optical resonator with a variable beamsplitter (VBS) of tunable coupling $\gamma(t)$. 
In memory mode, an incoming wavepacket (label ``1'') is absorbed and later released; in beamsplitter mode, the input (1) interferes with the intra-resonator mode (2), producing outputs (1',2'). 
\textbf{b} Conventional breeding protocol requiring many parallel non-Gaussian sources and iterative beamsplitters. 
\textbf{c} Time-domain-multiplexed (TDM) breeding protocol using a single source and the dual-mode memory, sequentially implementing the equivalent circuit of \textbf{b}.}
\end{figure*}
Quantum information processing with light is one of the most promising avenues for achieving large-scale quantum computing. While large-scale Gaussian operations have already been demonstrated \cite{warit_cluster,mikkel_cluster,cluster_operation}, the true breakthroughs required for fault-tolerance \cite{gottesman_knill} and quantum supremacy \cite{qec_nogo} rely on the generation of non-Gaussian states, such as the Gottesman-Kitaev-Preskill (GKP) state \cite{gkp}. However, generating optical non-Gaussian states with significant non-Gaussianity is challenging due to the inherent small non-linearity in optics. Breeding protocols, which iteratively enhance non-Gaussianity by entangling multiple non-Gaussian states and conditioning on a part of the modes, have emerged as a potential solution. Various types of breeding protocols have been proposed \cite{cat_breeding_theory,gkp_breeding,photon_catalysis} and experimentally demonstrated \cite{cat_breeding_exp,konno_gkp}, showing enhanced non-Gaussianity from initially weakly non-Gaussian states. However, scaling up these protocols --- critical for practical quantum computing --- is hindered by the probabilistic nature of heralded state generation, which introduces randomness and unpredictability in the timing of state generation.

To address these challenges, we propose and demonstrate a protocol for a scalable non-Gaussian state generation using an all-optical, continuous-time resonator quantum memory, in a time-domain-multiplexed (TDM) manner. By mitigating the timing randomness of state generation through the use of a quantum memory, our approach enables a large-scale breeding circuits, equivalent to Fig.~\ref{fig:dual_mode_breeding}b, using only a single state generator and a single quantum memory (Fig.~\ref{fig:dual_mode_breeding}c). In a proof-of-principle experiment, we successfully demonstrated both cat \cite{cat_breeding_theory,cat_breeding_exp} and GKP \cite{gkp_breeding,konno_gkp} breeding, and generated an approximate GKP state with two Wigner-negative regions from a single source of squeezed single-photon states, showcasing the feasibility of our approach.

The core innovation lies in the dual-mode operation of the resonator memory: the ``memory mode'' stores quantum states in the intra-resonator mode, while the ``entangling mode'' implements beamsplitter gates between incoming wavepackets and the intra-resonator mode in a TDM manner. This dual functionality eliminates timing randomness and facilitates scalable, compact entangling operations.

We further demonstrate, for the first time, the full operation of an optical resonator memory—encompassing writing, storage, and readout—using squeezed single-photon states with varying squeezing levels. The measured memory lifetime and dephasing time are $T_1 = \SI{2.3}{\micro s}$ and $T_\phi = \SI{0.96}{\micro s}$, respectively.

These advances provide a scalable pathway for generating complex non-Gaussian states critical for quantum information processing. Beyond its significance for optical quantum computing, our protocol has broad implications, potentially advancing quantum communication networks, quantum sensing, and metrology.
\begin{figure*}[htbp]
    \centering
    \includegraphics[scale=1]{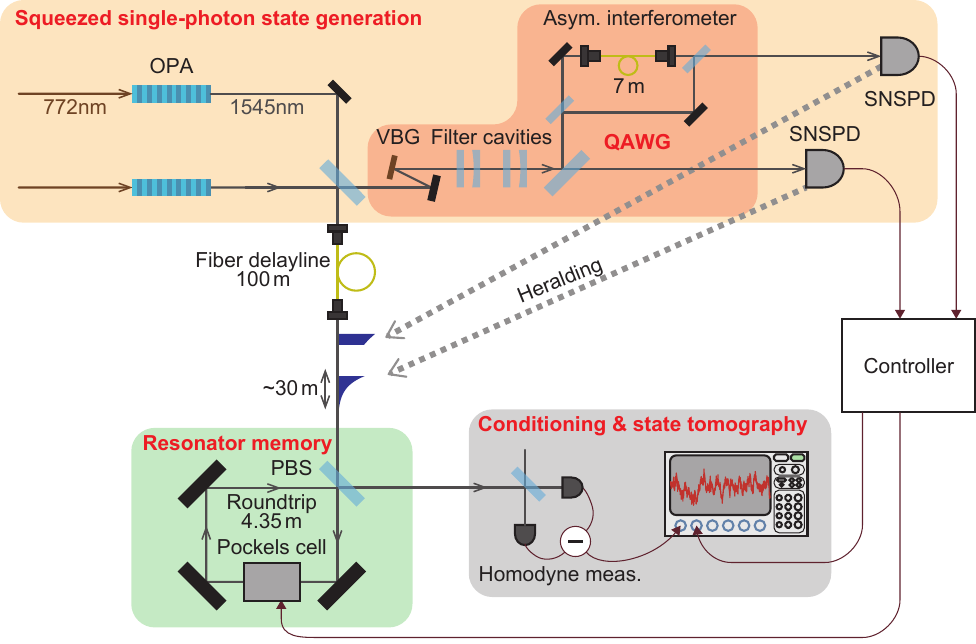}
    \caption{\bodyfigurelabel{fig:exp-setup} 
    Detailed experimental setup. OPA: optical parametric amplifier; VBG: volume Bragg grating; PBS: polarization beam splitter; QAWG: quantum arbitrary waveform generator~\cite{QAWG}; SNSPD: superconducting nanowire single-photon detector.}
\end{figure*}
\section{Dual-mode operation of resonator memory}
Our quantum memory can be modeled as an optical resonator with a variable output coupling constant $\gamma(t)$, physically implemented using a Pockels cell and a polarizing beamsplitter (PBS). See Sec.~\ref{sec:cavity_detail} for the detail. By adjusting $\gamma(t)$, the system can switch between two operational modes: memory mode and entangling mode (Fig.~\ref{fig:dual_mode_breeding}a). In memory mode, the memory can absorb quantum states with arbitrary wavepacket shapes, store them for a specified time, and release them at continuously arbitrary times in an arbitrary wavepacket. In entangling mode, the system implements a beamsplitter gate with a variable transmittance between the intra-resonator mode and an incoming wavepacket with an arbitrary shape. One output state remains in the resonator, while the other is emitted as a wavepacket with a shape determined by the input. 

A unique aspect of our system is its ability to store quantum states in continuous time, unlike conventional loop-based architectures \cite{all_optical_memory,loop_breeding,takeda_loop,kaneda_loop}, which are inherently limited to discrete-time operation. This capability arises from operating in the regime where the wavepacket duration is much longer than the resonator roundtrip time, so the state is stored in a delocalized resonance mode rather than merely as a circulating pulse. In this regime, storage and readout are implemented through resonant interference effects. In the frequency domain, this corresponds to storing information in a single resonance mode, whereas loop-based memories are intrinsically multimode, relying on multiple resonances. This feature of our memory offers a natural compatibility with continuous-wave state generation, which typically achieves higher efficiency \cite{pure_photon} and rates \cite{fast_cat} compared to pulsed regimes. Furthermore, this approach aligns naturally with the large-scale cluster states demonstrated in previous works \cite{warit_cluster,mikkel_cluster,cluster_operation}, which were also implemented in the continuous-wave regime.

\textit{Memory mode:}
In memory mode, the quantum memory stores an input quantum state with an arbitrary real temporal mode function, $g_{\text{in}}(t)$. During the writing process, the coupling is dynamically adjusted according to the following equation:
\begin{align}
    \gamma(t) = \frac{g_{\text{in}}(t)^2}{\int_{-\infty}^t g_{\text{in}}(t')^2 \, dt'}.
\end{align}

While the state is stored, the coupling is set to zero ($\gamma(t) = 0$). The readout process retrieves the state in an arbitrary temporal mode $g_{\text{out}}(t)$ by applying the following coupling:
\begin{align}
    \gamma(t) = \frac{g_{\text{out}}(t)^2}{\int_t^{\infty} g_{\text{out}}(t')^2 \, dt'}.
\end{align}

\textit{Entangling mode:}
In entangling mode, the stored quantum state interacts with the input wavepacket, $g_{\text{in}}(t)$, via a dynamically controlled coupling, $\gamma(t)$, given by:
\begin{align}
    \gamma(t) = \frac{g_{\text{in}}(t)^2}{\frac{T_f}{1-T_f} + \int_{-\infty}^t g_{\text{in}}(t')^2 \, dt'}.
\end{align}
This operation effectively implements a beamsplitter gate, where $T_f$ represents the transmittance of the gate. One of the output states is stored in the memory, while the other is emitted as a wavepacket $g_{\text{out}}(t)$, whose form determined by $g_{\text{in}}(t)$ and $T_f$.

The detail of the derivation of operational principles of the dual-mode resonator memory is given in Appendix~\ref{sec:derivation_cavity_memory}.

\section{Wavepacket engineering}
Our scheme enables memory and entangling operations for arbitrary shapes of input wavepackets by tailoring the control pulses applied to the Pockels cell. This capability is further enhanced by combining it with the quantum arbitrary waveform generator (QAWG) technique \cite{QAWG}, which allows precise control over the wavepacket shape of the input wavepacket via optical filtering of the heralding path. By engineering the wavepacket shape, the control pulses can be designed to satisfy specific experimental constraints.

In particular, rectangular control pulses taking two discrete values, $\gamma(t) \in \{0, \gamma_0\}$, can be implemented using a fast electric switch and a high-voltage source. For the memory writing operation, an exponentially rising wavepacket can be employed (Fig.~\ref{fig:memory-performance}b):
\begin{align}
    g_{\mathrm{in}}(t) \propto 
    \begin{cases}
        e^{\frac{\gamma_0}{2} t}, & t \leq 0, \\
        0, & t > 0,
    \end{cases}\label{eq:exp_rising}
\end{align}
corresponding to the control pulse:
\begin{align}
    \gamma(t) =
    \begin{cases}
        \gamma_0, & t \leq 0, \\
        0, & t > 0.
    \end{cases}\label{eq:pulse_exp_rising}
\end{align}

The idea of QAWG is that, given sufficiently broadband entanglement before detection, the wavepacket shape in the signal mode is the time-reversal of the impulse response of the filtering function in the idler mode. Hence, the exponentially rising wavepacket in Eq.~\eqref{eq:exp_rising} can be realized using a Lorentzian filter from a cavity (see Sec.~\ref{sec:state_generation_method}).

For the memory readout operation, the control pulse is given by:
\begin{align}
    \gamma(t) =
    \begin{cases}
        \gamma_0, & t \geq 0, \\
        0, & t < 0,
    \end{cases}\label{eq:pulse_exp_decaying}
\end{align}
resulting in an exponentially decaying wavepacket:
\begin{align}
    g_{\mathrm{out}}(t) \propto 
    \begin{cases}
        e^{-\frac{\gamma_0}{2} t}, & t \geq 0, \\
        0, & t < 0.
    \end{cases}\label{eq:exp_decaying}
\end{align}

For the entangling operation, a time-bin wavepacket can be employed (Fig.~\ref{fig:breeding_result}b):
\begin{align}
    g_{\mathrm{in}}(t) \propto 
    \begin{cases}
        e^{\frac{\gamma_0}{2} t}, & 0 \leq t \leq t_0, \\
        0, & \text{otherwise}.
    \end{cases}\label{eq:time_bin}
\end{align}
This wavepacket is generated by passing the idler through an asymmetric interferometer introducing a delay equal to the time-bin length~\cite{QAWG}. The corresponding control pulse and output wavepacket are
\begin{align}
    \gamma(t) =
    \begin{cases}
        \gamma_0, & 0 \leq t \leq t_0, \\
        0, & \text{otherwise},
    \end{cases}\label{eq:pulse_time_bin}
\end{align}
and
\begin{align}
    g_{\mathrm{out}}(t) \propto 
    \begin{cases}
        e^{-\frac{\gamma_0}{2} t}, & 0 \leq t \leq t_0, \\
        0, & \text{otherwise},
    \end{cases}
\end{align}
respectively.To generate both types of wavepackets from the same source, the idler path is split into two arms, each routed through a different path and detected by separate SNSPDs, enabling heralding of distinct wavepacket shapes.

Combining the generation of quantum states in these wavepackets using the QAWG technique, and the rectangular control electric pulses enabled by a fast electric switch, both the memory mode and entangle mode operations can be performed in an experimentally feasible manner.

\section{TDM breeding protocol}
By combining the memory and entangling operations of the resonator memory (Fig.~\ref{fig:dual_mode_breeding}\textbf{b}), an arbitrary-step TDM breeding protocol can be implemented using a single state generator and a single resonator memory. Multiple wavepackets are sequentially stored in the memory, and the breeding protocol proceeds by alternating memory-mode and entangling-mode operations.

As examples, we demonstrate both the cat breeding \cite{cat_breeding_theory,cat_breeding_exp} and GKP breeding \cite{gkp_breeding,konno_gkp} protocols, realized by $p$- and $x$-quadrature homodyne detection, respectively. Detailed derivations are given in Appendix~\ref{sec:tdm_breeding_detail}, which also includes a two-memory scheme for generating a GKP state from squeezed single-photon states.  

Both protocols use the cat state
\begin{align}
\frac{1}{\mathcal{N}}\big(\ket{i\alpha} \pm \ket{-i\alpha}\big)
\end{align}
as input.  
In cat breeding, the output mode after the beamsplitter gate is projected onto $p=0$. The state in the memory after the $k$-th step is
\begin{align}
\ket{i\sqrt{k}\alpha} \pm \ket{-i\sqrt{k}\alpha},
\end{align}
with amplitude increased to $\sqrt{k}\alpha$.

In GKP breeding, $x$-quadrature projection ($x=0$) is used instead, producing a GKP state with $k{+}1$ peaks in $p$:
\begin{align}
\sum_{m=0}^{k+1}\mqty(k+1\\m)s^m \ket{i(-k-1+2m)/\sqrt{k}\alpha}.
\end{align}
In our implementation, a squeezed single-photon state is used as a good approximation of a small-amplitude cat state~\cite{squeezed_sp_cat}. See Appendix~\ref{sec:tdm_breeding_detail} for further details.

\section{Methods}
\subsection{Experimental Setup}
The full experimental setup is illustrated in Fig.~\ref{fig:exp-setup}. A continuous-wave laser with a wavelength of \SI{1545.32}{nm} is used to generate squeezed single-photon states, which are subsequently stored in the quantum memory. The output of the quantum memory is measured via a single balanced homodyne detector, which is used for both conditioning and state tomography. The details of each component are described below. See Appendix \ref{sec:control_seq} for the details of the control sequences used in the experiments.

\subsection{Generation of Squeezed Single-Photon States}\label{sec:state_generation_method}
Squeezed single-photon states are generated using the generalized photon subtraction method \cite{GPS,gps_exp}. Two squeezed vacuum states are produced in Type-0 periodically poled lithium niobate (PPLN) waveguides \cite{NTT_OPA}. These states are combined at a beamsplitter to entangle the signal and idler modes. The squeezing level and beamsplitter ratio are chosen to ensure a low mean photon number in the idler mode, with small multi-photon contributions \cite{gps_exp}.

The idler mode is detected using two superconducting nanostrip single-photon detectors (SNSPDs) made of NbTiN \cite{sspd}, which herald the generation of squeezed single-photon states in the signal mode. To produce different wavepacket shapes (exponential-rise and time-bin), the idler mode is processed using QAWG \cite{QAWG}. It passes through a volume Bragg grating (VBG) with a half-width half-maximum (HWHM) of \SI{5.6}{GHz}, followed by filter cavities with HWHMs of \SI{87}{MHz} and \SI{1.5}{MHz}. These filters reduce the idler photon frequency bandwidth to a single peak with an HWHM of \SI{1.5}{MHz}.

The filtered idler photon is then split and directed to two SNSPDs (SNSPD1 and SNSPD2). The paths to these SNSPDs are designed to herald different wavepackets in the signal mode. For the exponential-rise wavepacket, the idler photon is detected directly by SNSPD1. For the time-bin wavepacket, the idler photon traverses an asymmetric interferometer with a \SI{7}{m} fiber delay before detection by SNSPD2. The detection rates are approximately \SI{1.0e4}{} counts per second (cps) for SNSPD1 and \SI{1.5e3}{cps} for SNSPD2. The total efficiency of the signal path, including the detection efficiency is estimated to be \SI{70}{\%} from the measurement of generated state without passing through the quantum memory. 

\subsection{Resonator Quantum Memory}\label{sec:cavity_detail}
The quantum memory consists of an optical resonator equipped with a RbTiOPO$_4$ (RTP) Pockels cell and a polarizing beamsplitter (PBS) serving as the output coupler. The round-trip length of the resonator is $L = \SI{4.35}{m}$. The effective coupling, $\gamma(t)$, of the output coupler is controlled by applying a voltage $V(t)$ to the Pockels cell according to:
\begin{align}
    \gamma(t) = \frac{c}{L}\left[1 - \cos\left(\frac{\pi}{V_\pi}V(t)\right)\right],\label{eq:gamma_V_rel}
\end{align}
where $V_\pi = \SI{1.8e3}{V}$ is the half-wave voltage of the Pockels cell and $c$ is the speed of light. In the experiments, square voltage pulses are applied using a high-voltage switch, resulting in $\gamma(t) = \gamma_0 = 2\pi\times \SI{1.5}{MHz}$, matching the input wavepacket bandwidth. See Appendix \ref{sec:control_seq} for the details of the control sequences.

\subsection{Homodyne Measurement and Data Analysis}
A single homodyne detector with a bandwidth of \SI{200}{MHz} is used for both conditioning in the breeding protocol and quantum state tomography. For memory performance characterization (Fig.~\ref{fig:memory-performance}), 10,000 frames of data were collected for each of six measurement phases (0, 30, 60, 90, 120, 150 degrees). The wavepackets of quantum states were estimated via principal component analysis (PCA) \cite{pca}.

In the TDM breeding experiment (Fig.~\ref{fig:breeding_result}), a waveguide electro-optic phase modulator with a \SI{10}{GHz} bandwidth is used for rapidly switching the local oscillator phase. This switching occurred after measuring the first wavepacket, enabling quantum state tomography on the second wavepacket conditioned on the first. A total of \SI{2e5}{} data frames were recorded, with PCA performed separately for the first and second wavepackets. Conditioning ($|p| \leq 0.1$) on the first wavepacket resulted in \SI{2e4}{} conditioned events. These data were used for maximum likelihood state tomography \cite{maxlike}.

\begin{figure*}[htbp]
    \centering
    \includegraphics[scale=1]{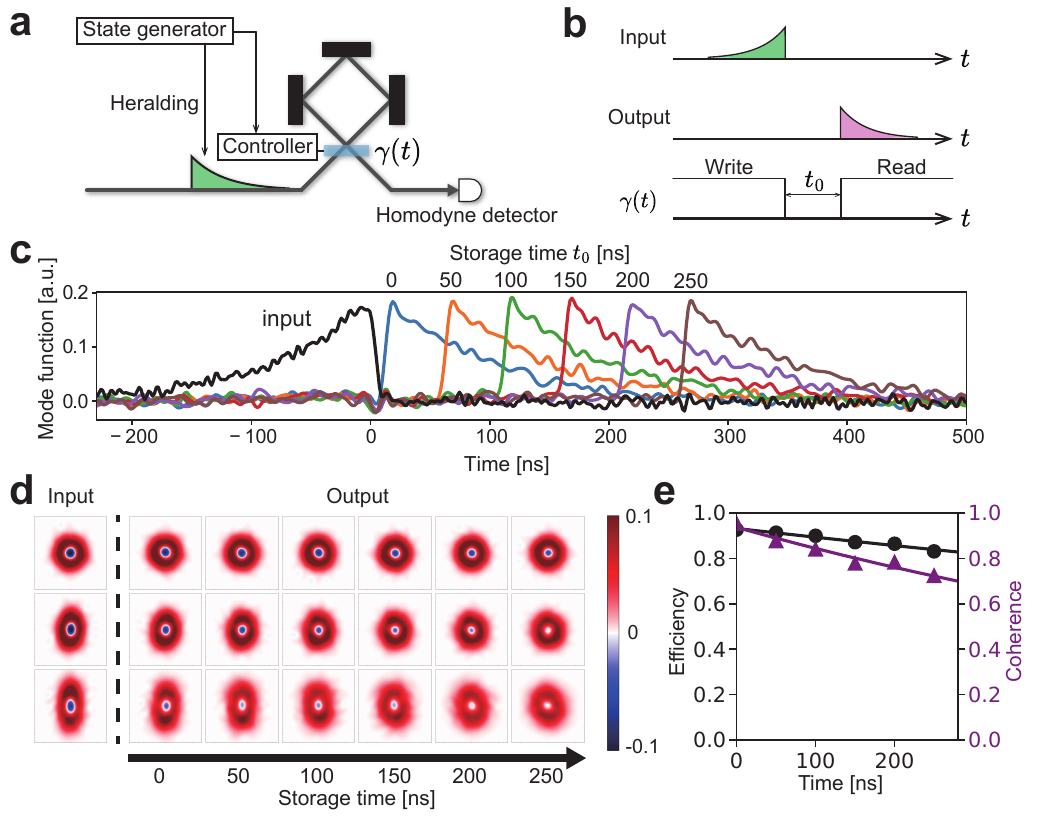}
    \caption{\bodyfigurelabel{fig:memory-performance} 
    Performance of the resonator memory. \textbf{a} Schematic of the experimental setup. \textbf{b} Pulse sequence applied during the experiment. \textbf{c} Temporal-mode functions of the input and output wavepackets for different storage times $t_0$. \textbf{d} Wigner functions of the input and output states for varying squeezing levels and storage times (axis range $x,p \in [-5,5]$, with $[\hat{x},\hat{p}] = i$). \textbf{e} Memory efficiency and coherence, obtained from storing a single-photon state (first row of \textbf{d}) and a squeezed single-photon state (last row of \textbf{d}), respectively; the solid lines are exponential fits, yielding a total efficiency of \SI{93}{\%}, an energy relaxation time $T_1 = \SI{2.3}{\micro s}$, and a pure dephasing time $T_\phi = \SI{0.96}{\micro s}$.}
\end{figure*}

\begin{figure*}[htbp]
    \centering
    \includegraphics[scale=1]{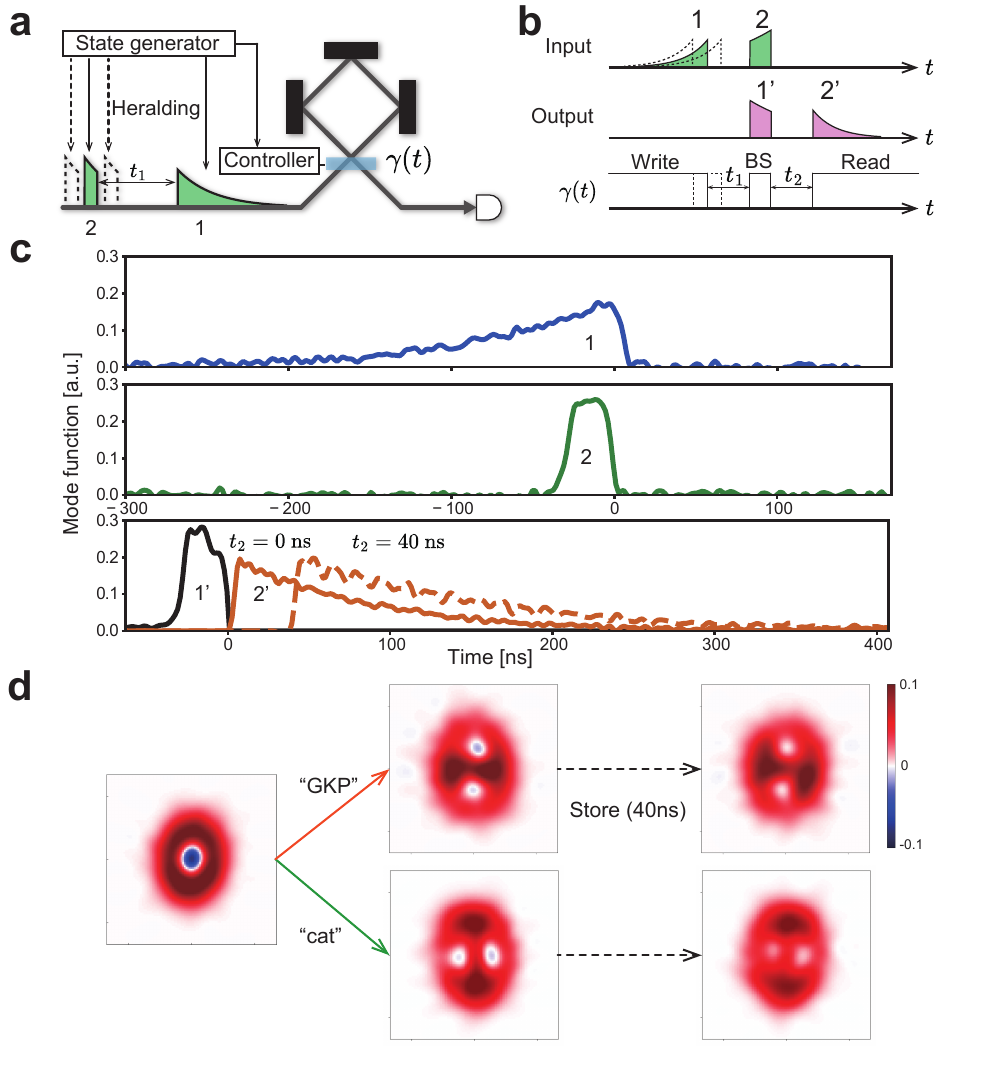}
    \caption{\bodyfigurelabel{fig:breeding_result} Implementation of the TDM breeding protocol using the resonator memory. \textbf{a} Schematic of the experimental setup. \textbf{b} Pulse sequence applied during the experiment. \textbf{c} Experimentally obtained temporal-mode functions: (first and second rows) input wavepackets, and (third row) output wavepackets. The second output wavepackets are shown for two different storage times, $t_2 = 0$ and $t_2 = \SI{40}{ns}$. \textbf{d} Wigner functions of (left) the input squeezed cat state, (middle) output state of the breeding protocol with $t_2 = 0$, and (right) output state with $t_2 = \SI{40}{ns}$. Results for both cat and GKP breeding are presented.}
\end{figure*}

\section{Results}
\subsection{Storage of squeezed single-photon states}
We first benchmarked the performance of our quantum memory for the memory operation. A squeezed single-photon state was generated via the generalized photon subtraction (GPS) method \cite{GPS}, and was input into the memory (Fig.~\ref{fig:memory-performance}a). The state was stored for time $t_0$ and then was readout and measured using homodyne detection (Fig.~\ref{fig:memory-performance}b). Figure \ref{fig:memory-performance}c shows the input and output wavepackets experimentally obtained by performing PCA of the homodyne signals for different storage times. The reconstructed Wigner function of the measured state without loss correction is shown in Fig.~\ref{fig:memory-performance}d. Note that we use the convention $\hbar=1$ throughout this paper.

From measurements with an (unsqueezed) single-photon input, we estimated the total memory efficiency to be \SI{93}{\%} and the energy relaxation time $T_1 = \SI{2.3}{\micro s}$ by fitting the decay of the single-photon component. For the squeezed single-photon input, the pure dephasing time $T_\phi = \SI{0.96}{\micro s}$ was extracted from the decay of the normalized coherence, obtained from the off-diagonal density matrix elements. Appendix~\ref{sec:imperfection} provides the detailed definitions of these quantities and presents simulations of the Wigner function evolution using the extracted parameters.

This marks the first demonstration of a fully functional resonator-based memory, including the complete process of writing to the memory, where previously, only storage and readout had been shown, with quantum memory directly integrated with the state generator \cite{memory_hashimoto}.

\subsection{TDM breeding}
Next, we implemented the TDM breeding protocol between two squeezed single-photon states from the same source (Figs.~\ref{fig:breeding_result}a and \ref{fig:breeding_result}b). These squeezed single-photon states can be treated as good approximations of small-amplitude squeezed cat states \cite{gps_exp}. The input wavepackets were shaped using QAWG \cite{QAWG} into exponentially-rising (Eq.~(\ref{eq:exp_rising})) and time-bin (Eq.~(\ref{eq:time_bin})) forms. These choices correspond to rectangular control pulses (Eqs.~(\ref{eq:pulse_exp_rising}) and (\ref{eq:pulse_time_bin})), as shown in Fig.~\ref{fig:breeding_result}b. Figure \ref{fig:breeding_result}c shows the experimentally obtained input and output wavepackets. The first wavepacket was stored in the memory while awaiting the second heralding event. Here, the storage time $t_1$ varied randomly depending on the heralding timing, and the events satisfying $0<t_1<200$~\si{ns} were collected. Once the second wavepacket arrived, the memory was operated in the entangling mode, emitting one of the output wavepackets. After the second storage period $t_2$, the intra-resonator mode was released and measured. We performed quantum state tomography on the state of the second wavepacket, conditioned on the first wavepacket by homodyne measurement. Depending on the measurement basis, both cat-breeding and GKP-breeding protocols were implemented.

We first characterized the generated quantum state immediately after the breeding operation ($t_2 = 0$). In both the GKP and cat-breeding cases, two Wigner-function negativities were observed without any loss correction, indicating the non-classicality of the states. We also investigated the storage of the bred states for $t_2 = 40$~ns. While the Wigner negativity was almost lost after this delay, the reconstructed Wigner function still retained the qualitative features of the target states. Simulations of the Wigner function evolution using the measured memory parameters are provided in Appendix~\ref{sec:imperfection}. For the GKP state with $t_2=0$, and defining the stabilizers $\hat{S}_x = e^{i g \hat{x}}$ and $\hat{S}_p = e^{2\pi i \hat{p}/g}$, we measured expectation values $(|\langle S_x \rangle|, |\langle S_p \rangle|) = (0.14, 0.23)$ with $g = 2.46$.

The successful interference rate was \SI{20}{}~counts per second (cps). We define a dimensionless parameter $k$ as the storage time normalized by the wavepacket duration. As discussed in Appendix~\ref{sec:rate_detail}, $k$ determines the scalability of the breeding protocol. In our experiment, we obtain $k = 3.8$, representing a $10^2$-fold enhancement compared with $k = 0.03$ in the previous GKP-breeding experiment without quantum memory~\cite{konno_gkp}.

\section{Conclusion \& Discussions}
We have proposed and experimentally demonstrated a novel and scalable scheme for generating optical non-Gaussian states using a resonator quantum memory equipped with dual-mode operation. This architecture, capable of both storage (“memory mode”) and beamsplitter interactions (“entangling mode”), enables time-domain-multiplexed cat- and GKP-state breeding protocols using only a single non-Gaussian state source. In our proof-of-principle experiment, we successfully generated Wigner-negative states from a single source. The dual-mode functionality of the memory eliminates timing randomness in heralded state generation and provides a compact, resource-efficient platform for scaling up optical quantum information processing. The quality of the obtained GKP state, $(|\langle S_x \rangle|, |\langle S_p \rangle|) = (0.14, 0.23)$, is slightly lower than the state-of-the-art in optics ($(0.17,0.22)$ for breeding~\cite{konno_gkp} and $(0.24,0.27)$ for multi-photon measurement \cite{xanadu_gkp}) and significantly lower than results from other physical platforms, such as motional modes of trapped ions ($(0.41,0.56)$~\cite{ion_gkp}) and superconducting resonators ($(0.88,0.94)$ \cite{ecd_gate}). Further improvements in the quantum memory's performance and extension to multi-step breeding will enhance the quality and scalability of the generated states.

The experimental demonstration reported here represents the first complete operation of an optical resonator memory—including writing, storage, and readout—using squeezed single-photon states with varying squeezing levels. The measured lifetime and dephasing time, $T_1 = \SI{2.3}{\micro s}$ and $T_\phi = \SI{0.96}{\micro s}$ respectively, confirm the practicality of our approach and its suitability for more sophisticated non-Gaussian state generation protocols. 

We have shown that the interference parameter $k$—quantifying the rate of successful interference events—achieves a 100-fold enhancement compared to the memoryless case. Nevertheless, the actual interference rate improves by only a factor of two, as the bandwidth of state generation is constrained by the narrow operational bandwidth ($\sim$\SI{10}{MHz}) of the Pockels cell driver. This restriction reduces the generation rate of squeezed single-photon sources and increases the wavepacket duration, limiting this work to a single breeding step. This is not a fundamental limitation and could be overcome by employing faster optical switches, such as integrated-optics Mach–Zehnder interferometers~\cite{MZI_memory,integrated_optics} or all-optical switching~\cite{all_optical_switch}, together with high-rate, broadband non-Gaussian state generation~\cite{fast_cat} and multi-photon detection~\cite{multi_photon_endo}.

A particularly promising aspect of our quantum memory is its potential to handle arbitrary input wavepackets. The ability to entangle wavepackets with different shapes, as demonstrated in our current results, underscores this potential. Future refinements in controllability beyond simple rectangular pulses could enable the memory to fully manage arbitrary wavepacket shapes, addressing challenges such as mode-matching across multiple quantum sources, including quantum dots \cite{quantum_dot}. This capability would significantly enhance the platform's versatility and broaden its range of applications.

The insights from our work are not limited to the demonstrated system but can inform the development of other types of quantum memories and extend to a variety of physical platforms. As we continue to refine these technologies, the ability to generate and manipulate complex quantum states on demand will be a cornerstone for unlocking the full potential of quantum computing, communication, and sensing. This research represents a critical step toward scalable, fault-tolerant quantum technologies and the realization of universal quantum systems.

\begin{acknowledgments}
This work was partially supported by JST Moonshot R\&D (Grant Nos.~JPMJMS2064, JPMJMS2066), JST PRESTO (Grant No.~JPMJPR2254), JSPS KAKENHI (Grant Nos.~23KJ0498, 24K01374), the UTokyo Foundation, and donations from Nichia Corporation.
\end{acknowledgments}




\bibliography{main}
\newpage
\appendix
\section{Theory of resonator quantum memory}\label{sec:derivation_cavity_memory}
We describe the dual-mode operation of our resonator quantum memory using input–output theory. Let $a$, $b_{\text{in}}$, and $b_{\text{out}}$ denote the mode operators for the resonator, input field, and output field, respectively. The power transmittance of the output coupler, $T(t)$, is made time-dependent by controlling the voltage applied to the Pockels cell. Assuming $T(t) \ll 1$ and applying the rotating-wave approximation, the resonator can be modeled as a single-mode oscillator. The out-coupling rate is expressed as $\gamma(t) = \frac{c}{L}T(t)$, where $c$ is the speed of light and $L$ is the round-trip length of the resonator. The time evolution of $a(t)$ is governed by the Langevin equation:
\begin{align}
    \dot{a} = -\frac{\gamma(t)}{2} a + \sqrt{\gamma(t)} b_{\text{in}}(t).
\end{align}

The solution to this equation is:
\begin{align}
    a(t) = F(t)a(t_0) + \int_{t_0}^t \frac{F(t)}{F(t')} \sqrt{\gamma(t')} b_{\text{in}}(t') dt,
\end{align}
where $F(t) = \exp\left(-\int_{t_0}^t \frac{\gamma(t')}{2} dt'\right)$.

By symmetry, reversing time in the Langevin equation gives:
\begin{align}
    \dot{a} = \frac{\gamma(t)}{2} a + \sqrt{\gamma(t)} b_{\text{out}}(t),
\end{align}
with the corresponding solution:
\begin{align}
    a(t) = \frac{1}{F(t)} a(t_0) + \int_{t_0}^t \frac{F(t')}{F(t)} \sqrt{\gamma(t')} b_{\text{out}}(t') dt.
\end{align}

Defining $T_f = F(t_f)^2$, we introduce the input wavefunction $g_{\text{in}}(t)$ and its associated operator $B_{\text{in}}$ as:
\begin{align}
    g_{\text{in}}(t) &= \sqrt{\frac{T_f}{1-T_f}} \frac{\sqrt{\gamma(t)}}{F(t)}\\,B_{\text{in}} &= \int_{t_0}^{t_f} g_{\text{in}}(t) b_{\text{in}}(t) dt.
\end{align}
Similarly, the output wavefunction $g_{\text{out}}(t)$ and its corresponding operator $B_{\text{out}}$ are defined as:
\begin{align}
    g_{\text{out}}(t) &= \frac{1}{\sqrt{1-T_f}} F(t) \sqrt{\gamma(t)},\\ B_{\text{out}} &= \int_{t_0}^{t_f} g_{\text{out}}(t) b_{\text{out}}(t) dt.
\end{align}
These satisfy the normalization conditions:
\begin{align}
    \int_{t_0}^{t_f} g_{\text{in}}(t)^2 dt = \int_{t_0}^{t_f} g_{\text{out}}(t)^2 dt = 1,\\ [B_{\text{in}}, B_{\text{in}}^\dagger] = [B_{\text{out}}, B_{\text{out}}^\dagger] = 1.
\end{align}

Using the solutions above, we find the following relations:
\begin{align}
    a(t) &= \sqrt{T_f} a(t_0) + \sqrt{1-T_f} B_{\text{in}},\\B_{\text{out}} &= -\sqrt{1-T_f} a(t_0) + \sqrt{T_f} B_{\text{in}}.
\end{align}
This shows that the resonator effectively acts as a beamsplitter, coherently mixing the intra-resonator mode $a(t_0)$ with the input mode $B_{\text{in}}$ to produce the output mode $B_{\text{out}}$. The transmittance of this effective beamsplitter is $T_f$.

Conversely, for a given real temporal-mode function $g_{\text{in}}(t)$, we find:
\begin{align}
    \qty(\frac{1}{F(t)^2})' = \frac{1-T_f}{T_f} g_{\text{in}}(t)^2,
\end{align}
which leads to:
\begin{align}
    \frac{1}{F(t)^2} = 1 + \frac{1-T_f}{T_f} \int_{t_0}^{t} g_{\text{in}}(t')^2 dt'.
\end{align}
From this, we derive a simple expression for $\gamma(t)$ to realize a beamsplitter for the wavepacket $g_{\text{in}}(t)$:
\begin{align}
    \gamma(t) &= \qty[\log\left(1 + \frac{1-T_f}{T_f} \int_{t_0}^{t} g_{\text{in}}(t')^2 dt'\right)]' \\
    &= \frac{g_{\text{in}}(t)^2}{\frac{T_f}{1-T_f} + \int_{t_0}^{t} g_{\text{in}}(t')^2 dt'}.
\end{align}

When $T_f = 0$, the input quantum state is fully transferred to the intra-resonator mode, corresponding to the memory-mode operation. The condition for $\gamma(t)$ in this case is:
\begin{align}
    \gamma(t) = \frac{g_{\text{in}}(t)^2}{\int_{t_0}^{t} g_{\text{in}}(t')^2 dt'}.
\end{align}
By controlling $\gamma(t)$ according to this relation, one can implement a memory for an arbitrary input wavepacket $g_{\text{in}}(t)$. For the output wavepacket $g_{\text{out}}(t)$, similarly because
\begin{align}
    \qty(-F(t)^2)' = (1-T_f) g_{\text{out}}(t)^2,
\end{align}
we have
\begin{align}
    F(t)^2 = 1 - (1-T_f) \int_{t_0}^t g_{\text{out}}(t')^2 dt'.
\end{align}
By defining
\begin{align}
    G_{\text{in}}(t) &= \int_{t_0}^{t} g_{\text{in}}(t')^2 dt', \\
    G_{\text{out}}(t) &= \int_{t_0}^{t} g_{\text{out}}(t')^2 dt',
\end{align}
we obtain an explicit relation between $g_{\text{in}}(t)$ and $g_{\text{out}}(t)$:
\begin{align}
    \qty(1 - \frac{1-T_f}{T_f} G_{\text{in}}(t)) \qty(1 - (1-T_f) G_{\text{out}}(t)) = 1.
\end{align}

Finally, we remark that loop-based quantum memories \cite{all_optical_memory,loop_breeding,takeda_loop,kaneda_loop} superficially resemble our system but differ fundamentally in both spectral and temporal structure. In the frequency domain, the resonator memory stores information in a single resonance, while a loop memory is inherently multimode, supporting multiple resonances. In the time domain, a loop memory stores a pulse circulating in the loop, whereas a resonator memory stores a delocalized intra-resonator field. Our system operates in the regime where the wavepacket length ($\sim\SI{30}{m}$ in our case) exceeds the round-trip length (\SI{4.35}{m}), whereas loop memories operate in the opposite limit. This single-mode description breaks down when $T(t)$ becomes too large, making multimode effects non-negligible. We discuss this effect in Appendix~\ref{sec:multi_mode_effect}. A key advantage of the resonator memory is the ability to store and retrieve quantum states at arbitrary continuous times, in contrast to the discrete-time operation of loop memories.

\section{Theory of TDM breeding protocols}\label{sec:tdm_breeding_detail}
\begin{figure*}[htbp]
    \includegraphics[scale=1]{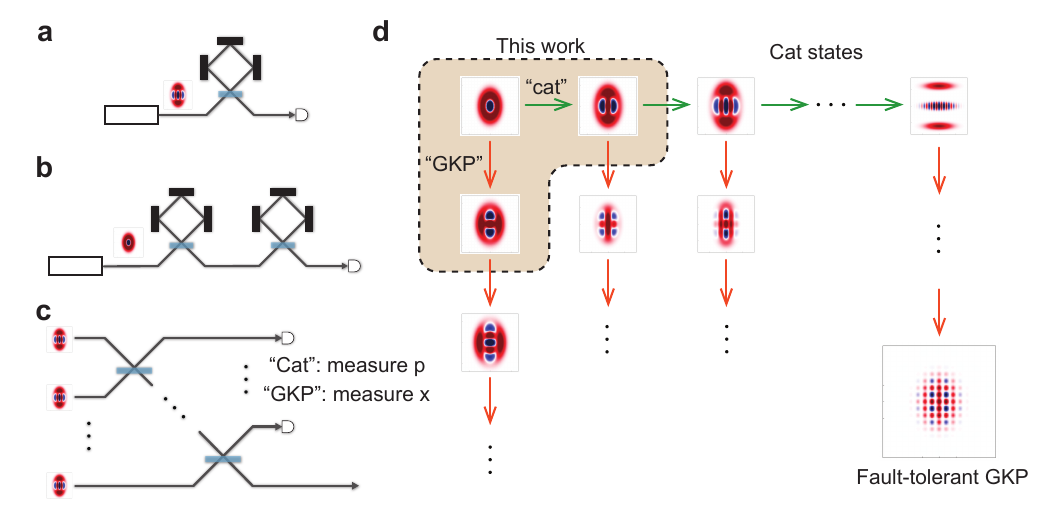}
    \caption{Implementation of the TDM GKP breeding protocol using resonator quantum memories. 
\textbf{a} Configuration with a single quantum memory, as implemented in the main text. 
\textbf{b} Configuration with two quantum memories. 
\textbf{c} Schematic representation of a protocol equivalent to that implemented by the system in \textbf{a}. 
\textbf{d} Evolution of the Wigner function during the breeding protocols.}
\label{fig:breeding_theory}

\end{figure*}

For non-Gaussian state generation protocols based on photon-number measurement, the stellar rank of the generated state, which is an indicator of its non-Gaussianity, is upper-bounded by the total number of photons detected \cite{stellar_rank,gbs_nongauss}. Generating useful non-Gaussian states typically requires a high stellar rank. For example, fault-tolerant GKP states require at least 20 photons \cite{gkp_nonlinear_sqz,xanadu_gkp}. In practice, this number varies depending on the protocol, e.g., 42 in \cite{gaussian_breeding}, 24 in \cite{xanadu_gkp}, and significantly more for near-deterministic schemes with higher success probability \cite{xanadu_architecture,kan_gkp}. The breeding protocol addresses this requirement by distributing the stellar rank across multiple modes and combining ---``breeding''--- them via measurement and conditioning.

Figure \ref{fig:breeding_theory}\textbf{c} shows the ``stair-case'' form of breeding circuit, which is equivalent to the time-domain multiplexed (TDM) implementation in Fig.~\ref{fig:breeding_theory}\textbf{a}, as realized in this work. By choosing the measurement basis ($x$ or $p$), this architecture can implement both ``GKP'' breeding, which generates a GKP state from cat states \cite{gkp_breeding,konno_gkp}, and ``cat'' breeding, which increases the amplitude of a cat state from multiple smaller cat states \cite{cat_breeding_theory,cat_breeding_exp}.

The detailed derivation of the breeding protocol is given in the Supplemental Material of Ref.~\cite{xanadu_architecture}; here we provide an overview in the context of our TDM structure.

\subsection{Cat breeding}
We consider the initial cat state
\begin{align}
\frac{1}{\mathcal{N}}\big(\ket{i\alpha} + s\ket{-i\alpha}\big),
\end{align}
where $\alpha > 0$ is the amplitude, $s \in {\pm 1}$ is the parity, and $\mathcal{N}$ is the normalization factor. At the $k$-th step, the states inside the memory and in the incoming wavepacket are:
\begin{align}
\ket{\psi}&=\frac{1}{\mathcal{N}_k}\qty(\ket{i\sqrt{k}\alpha}+s^k\ket{-i\sqrt{k}\alpha}),\\
\ket{\psi}_{\mathrm{in}}&=\frac{1}{\mathcal{N}}\qty(\ket{i\alpha}+s\ket{-i\alpha}).
\end{align}
Applying a beamsplitter with transmittance $T_k = k/(k+1)$ between these modes, followed by $p$-quadrature projection $\ket{p=0}$, yields:
\begin{align}
    \begin{split}    
    &\bra{p=0}\hat{\mathrm{BS}}(T_k)(\ket{\psi}\otimes \ket{\psi}_{\mathrm{in}})\\&\quad\quad\propto \ket{i\sqrt{k+1}\alpha}+s^{k+1} \ket{-i\sqrt{k+1}\alpha}
    \end{split}
\end{align}
producing a cat state with increased amplitude $\sqrt{k+1}\alpha$.
\subsection{GKP breeding}
For GKP breeding, the projection is onto $x=0$ instead of $p=0$. The state after the $k$-th step is:
\begin{align}
\ket{\psi}&=\frac{1}{\mathcal{N}_k}\sum_{m=0}^{k}\mqty(k\\m)s^m\ket{i(-k+2m)/\sqrt{k}\alpha},\\
\ket{\psi}_{\mathrm{in}}&=\frac{1}{\mathcal{N}}\qty(\ket{i\alpha}+s_1\ket{-i\alpha}).
\end{align}
After the conditioning, we obtain
\begin{align}
    \begin{split}    
    &\bra{x=0}\hat{\mathrm{BS}}(T_k)(\ket{\psi}\otimes \ket{\psi}_{\mathrm{in}})\\&\quad\quad\propto \sum_{m=0}^{k+1}\mqty(k+1\\m)s^m\ket{i(-k-1+2m)/\sqrt{k}\alpha}
    \end{split}
\end{align}
The number of peaks in the $p$-direction increases from $k+1$ to $k+2$.

Figure~\ref{fig:breeding_theory}\textbf{d} illustrates the evolution of Wigner functions in both protocols. Cat breeding increases the amplitude of the cat state, and the GKP breeding increases the number of peaks. A fault-tolerant GKP state can be obtained by starting from a sufficiently large cat state and applying multiple GKP breeding steps. In this work, both breeding types are demonstrated for small-amplitude cat states (Fig.~\ref{fig:breeding_result}\textbf{d} in the main text). Here note that a odd cat state with a small amplitude can be well-approximated by a squeezed single-photon state:
\begin{align}
    \frac{1}{\mathcal{N}}(\ket{i\alpha}-\ket{-i\alpha})\sim\hat{S}(r)\ket{1}
\end{align}
where $\hat{S}(r)=e^{(r/2)(\hat{a}^2-\hat{a}^{\dagger2})}$ and $r=-\cosh^{-1}(1/2+\sqrt{9+4\alpha^2}/6)^{1/2}$ \cite{squeezed_sp_cat}. 

Instead of preparing a large-amplitude cat source, two concatenated quantum memories can be used (Fig.~\ref{fig:breeding_theory}\textbf{b}): cat breeding is first performed in the initial memory to boost the amplitude, and the resulting larger cat state is then used as the input for GKP breeding in the second memory. This approach enables the generation of fault-tolerant GKP states using only a single small-cat source and two quantum memories.

\section{Discussion on interference rate in the breeding experiment}\label{sec:rate_detail}
\begin{figure*}[htbp]
\centering
\includegraphics[scale=1]{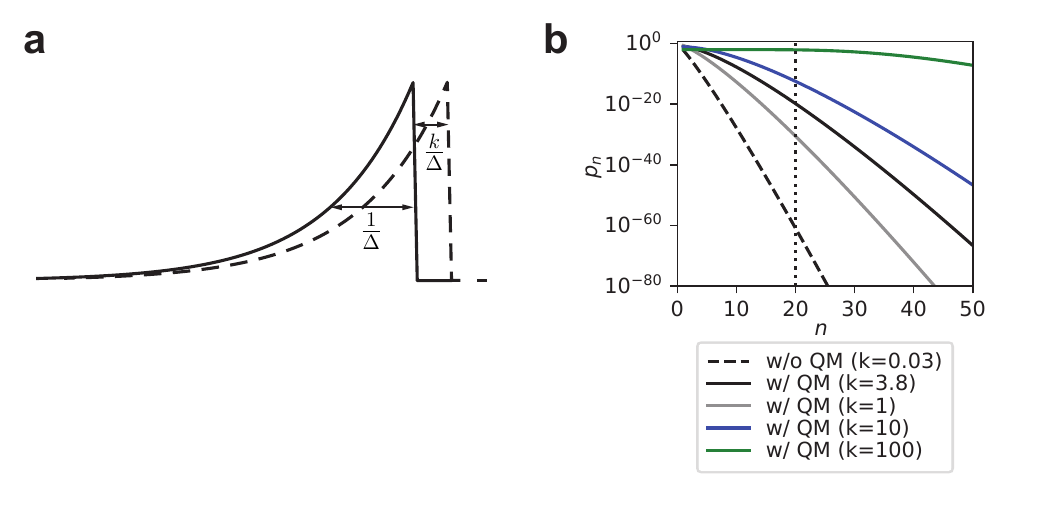}
\caption{\edfigurelabel{fig:wavepacket_modematch} \textbf{a} Illustration of interference of two wavepackets. $\Delta$: frequency bandwidth of the wavepacket, $k/\Delta$: the maximum time difference between two wavepackets for a successful interference. \textbf{b} Theoretical success probability $p_m$ for $m$ round breeding for different value of $k$ reflecting different storage time.}
\label{fig:interference_theory}
\end{figure*}
In this section, we explain the criteria for evaluating the interference rate in the breeding experiment. Let $r_0$ denote the heralding rate for generating a single quantum state, and let $\Delta$ represent the frequency bandwidth of the wavepackets, with $1/\Delta$ corresponding to their temporal width. Although the precise definition of $\Delta$ may vary depending on the wavepacket shape, we consistently define $\Delta$ as the full width at half maximum (FWHM) of the Lorentzian spectrum, which applies to the exponential wavepackets employed both in our experiment and in previous work~\cite{konno_gkp}. We assume Poissonian statistics for the heralding events, which is a reasonable approximation in the relevant regimes.

For two wavepackets to interfere successfully, they must overlap in time with sufficiently high fidelity. Let $k \ll 1$ be a dimensionless parameter characterizing the required degree of temporal mode matching. We post-select events such that two wavepackets arrive within a relative delay of $k/\Delta$. For each wavepacket, the probability of observing another heralding event within this temporal window is $r_0 k/\Delta$. Consequently, the expected rate of successful interference is given by
\begin{align}
r_{\text{BS}} = k \frac{r_0^2}{\Delta}.
\end{align}
The parameter $k$ reflects the requirement for temporal-mode overlap and is fundamentally limited in the absence of a quantum memory. With a quantum memory, however, the first quantum state can be stored until the second wavepacket arrives, effectively increasing $k$ beyond the restriction $k\ll1$ and thereby enhancing the interference rate.

In the previous GKP breeding experiment without quantum memory~\cite{konno_gkp}, the reported parameters were $r_0 = \SI{2e5}{cps}$, $\Delta = \SI{120}{MHz}$, and $r_{\text{BS}} = \SI{10}{cps}$, yielding $k = 0.03$. In our experiment, we have $r_0 = \SI{4e3}{cps}$, $\Delta = \SI{3}{MHz}$, and $r_{\text{BS}} = \SI{20}{cps}$, corresponding to $k = 3.8$, as noted in the main text.

From these values, the probability of detecting one photon within the temporal width of a single wavepacket is calculated as
\begin{align}
    p_1 = \frac{r_0}{\Delta}.
\end{align}
This $p_1$ corresponds to the heralding probability in the single-mode picture. In our experiment, we obtain $p_1 = \SI{1.3e-3}{}$. Theoretically, $p_1$ can be enhanced up to 0.25 using photon-number-resolving detectors~\cite{max_heralding,max_heralding_exp}. Currently, our $p_1$ is limited by the lack of photon-number-resolving capability and losses in the idler mode.

We now consider the scaling of the GKP breeding protocol initiated from squeezed single-photon states (see Appendix~\ref{sec:tdm_breeding_detail}). In an $(n-1)$-step breeding protocol involving $n$ wavepackets, at least $n$ heralding events must occur within a temporal window of $k/\Delta$. The success probability per wavepacket width is given by
\begin{align}
    p_n = \frac{1}{\max\{k,1\}} \left(1 - \sum_{n'=0}^{n-1} \frac{(k p_1)^{n'}}{n'!} e^{-k p_1} \right).
\end{align}
The corresponding success rate is $p_n \Delta$. 

Figure~\ref{fig:interference_theory}\textbf{b} plots $p_n$ as a function of $n$ for various values of $k$, assuming the ideal value $p_1 = 0.25$. The black dashed line ($k = 0.03$) corresponds to the previous experiment without quantum memory~\cite{konno_gkp}, while the black solid line ($k = 3.8$) reflects our present experiment. For the case $n = 20$, which is required for fault-tolerant GKP state generation \cite{gkp_nonlinear_sqz,xanadu_gkp}, our experimentally achieved $k$ yields a more than four orders of magnitude improvement in success probability compared to the case without a quantum memory. Although the resulting $p_{20}$ remains as low as $10^{-30}$, the curves for $k=1,10,100$ illustrate that further enhancement of $k$ could push the protocol into a practically viable regime even for large $n$.

We note that the probability shown here does not include the effect of homodyne conditioning, which is known to be near-deterministic~\cite{deterministic_breeding}. Additionally, the use of Poissonian statistics may introduce a constant-factor discrepancy. Finally, we remark that multi-photon detection schemes such as those in Refs.~\cite{xanadu_architecture,gaussian_breeding} fundamentally follow the same scaling as memoryless protocols, while schemes~\cite{xanadu_gkp} using pulsed inputs with intrinsically discretized timing correspond to $k = 1$ in the absence of quantum memory.

\section{Control Sequence for Pockels Cell}\label{sec:control_seq}
Figure \ref{fig:breeding_control} shows the control sequences used for each measurement discussed in the main text. The Pockels cell is equipped with two electrodes, each of which can assume one of two voltage levels, $0$ or $V_0$, depending on the states of the switches. The coupling constant $\gamma(t)$ is determined by the voltage difference $V(t)$ across the electrodes, as described by Eq.~(\ref{eq:gamma_V_rel}) in the methods section of the main text. In each sequence, the switches are controlled by a pulse generator (Quantum Composer 9528), which is triggered by heralding signals from single-photon nanowire detectors (SNSPDs).

\begin{figure*}[htbp]
\centering
\includegraphics[scale=1]{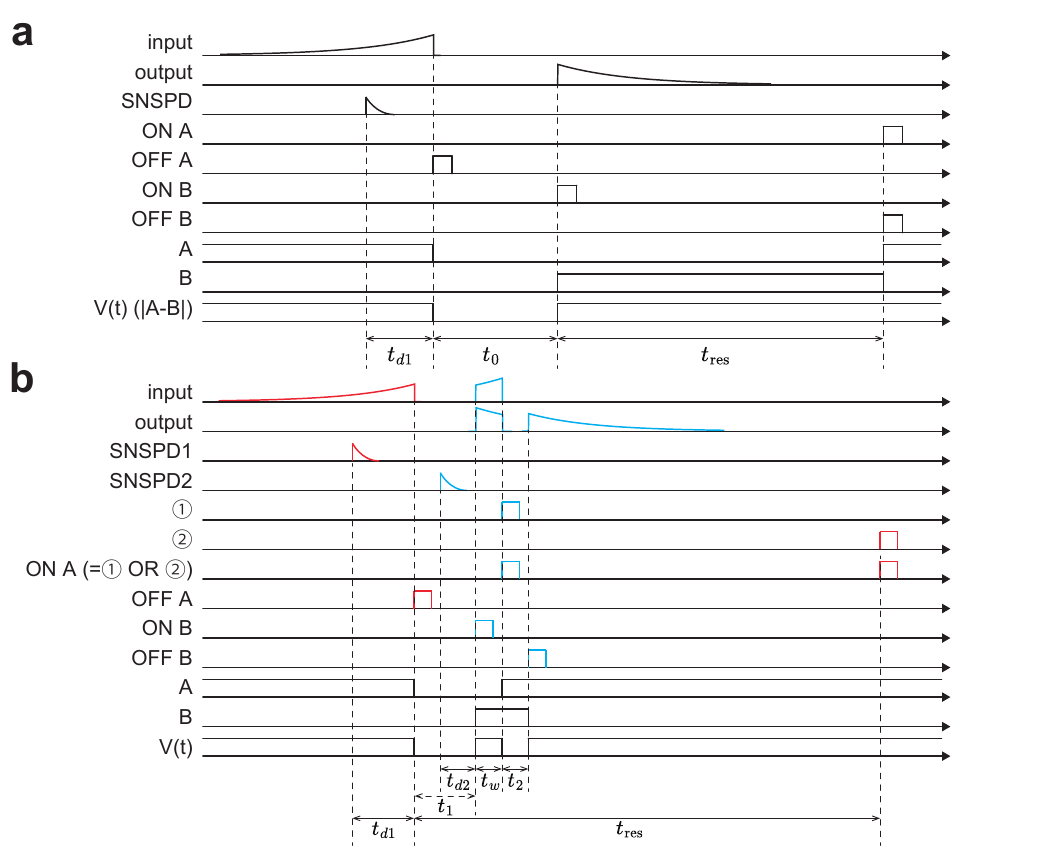}
\caption{\edfigurelabel{fig:breeding_control} \textbf{a} Control sequence of Pockels cell for the memory experiment (Fig. \ref{fig:memory-performance} in the main text). \textbf{b} Control sequence of Pockels cell for the TDM breeding experiment (Fig. \ref{fig:breeding_result} in the main text).}
\end{figure*}

\section{Discussion on the imperfection}\label{sec:imperfection}
We discuss here the main imperfections affecting the performance of the resonator quantum memory.
\subsection{Loss and dephasing}
During storage, the dominant errors are energy relaxation and pure dephasing. The evolution of a bosonic mode with energy relaxation time $T_1$ and pure dephasing time $T_\phi$ is described by the Lindblad master equation
\begin{equation}
\frac{d\rho}{dt} = \frac{1}{T_1} \, \mathcal{D}[a]\,\rho 
+ \frac{1}{T_\phi}\mathcal{D}[a^\dagger a]\,\rho
\end{equation}
where the dissipator is defined as:
\begin{equation}
\mathcal{D}[L]\,\rho = L \rho L^\dagger - \frac{1}{2} \left( L^\dagger L \rho + \rho L^\dagger L \right).
\end{equation}

In the Fock basis, the density matrix elements $\rho_{n,m}(t) = \bra{n} \rho(t) \ket{m}$ evolve as:
\begin{widetext}
\begin{equation}
\rho_{n,m}(t) = \sum_{k=0}^\infty \rho_{n+k,m+k}(0) 
\sqrt{\binom{n+k}{k} \binom{m+k}{k}} 
\left(1 - e^{-t/T_1}\right)^k 
e^{-\frac{t}{2T_1}(n + m)} 
e^{-t/T_\phi \cdot (n - m)^2}.\label{eq:dm_element}
\end{equation}
\end{widetext}

For the single-photon state input, the diagonal elements decay as:
\begin{equation}
\rho_{1,1}(t) \propto e^{- t / T_1}
\end{equation}
Thus, $T_1$ can be extracted experimentally by fitting this decay.

To estimate $T_\phi$, we focus on the $(n,m)=(0,2)$ element and define the normalized coherence
\begin{equation}
R(t) = \left(\frac{|\rho_{0,2}(t)|}{\sqrt{\rho_{0,0}(t) \rho_{2,2}(t)}}\right)^{1/4}
\end{equation}
This isolates the dephasing factor:
\begin{equation}
R(t) = R(0) \cdot e^{- \frac{t}{T_\phi}}.
\end{equation}
By fitting $R(t)$ to an exponential decay, we obtain $T_\phi$.

\begin{figure*}[htbp]
    \includegraphics[scale=1]{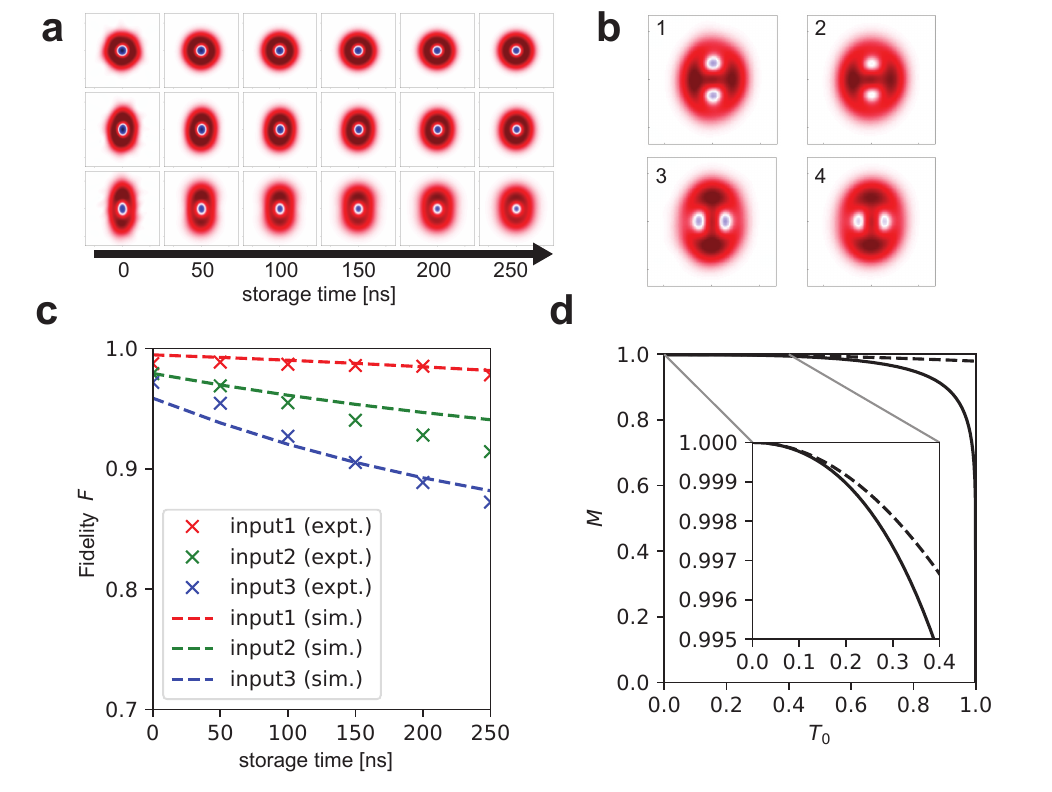}
    \caption{\textbf{a} Numerical simulation of storage of squeezed single-photon states, including loss and dephasing, corresponding to Fig.~\ref{fig:memory-performance}\textbf{d}. \textbf{b} Numerical simulation of the breeding protocol under the same imperfections, corresponding to Fig.~\ref{fig:breeding_result}\textbf{d}. \textbf{c} Fidelity between the retrieved and input states. Input 1, 2, 3 correspond to the upper, middle, and lower rows in both Fig.~\ref{fig:memory-performance}\textbf{d} and Fig.~\ref{fig:memory_sim}\textbf{a}. \textbf{d} Mode overlap $M$ between the readout wavepacket predicted by single-mode and multi-mode treatments, as a function of the output coupler transmittance $T_0$.}
    \label{fig:memory_sim}
\end{figure*}

In the experiment, we find $T_1 = \SI{2.3}{\micro s}$ and $T_\phi = \SI{0.96}{\micro s}$, as shown in Fig.~\ref{fig:memory-performance}\textbf{e}. Figure~\ref{fig:memory_sim}\textbf{a} shows numerical simulations of storage with these parameters, while Fig.~\ref{fig:memory_sim}\textbf{b} shows simulations of breeding protocols under the same conditions. Figure~\ref{fig:memory_sim}\textbf{c} compares the measured and simulated fidelity between the retrieved state and the input, showing good agreement.

\subsection{Multi-modeness}\label{sec:multi_mode_effect}

As described in Appendix~\ref{sec:derivation_cavity_memory}, the resonator quantum memory operates in a regime where it can be treated as a single-mode oscillator, enabling continuous-time writing and retrieval. Here, we quantify the validity of this approximation at finite finesse.

We consider the write/readout processes where $\gamma(t)\neq 0$. The dynamics of the storage mode can be written as
\begin{align}
    \dot{a}_0 = -\frac{\gamma(t)}{2} a_0 + \sqrt{\gamma(t)} b_{\text{in}}(t).
\end{align}
However, for low resonator finesse, the contributions from neighboring modes become non-negligible:
\begin{align}
    \dot{a}_{n} = -\frac{\gamma(t)}{2} a_n + \sqrt{\gamma(t)} b_{\text{in}}(t) +i n\Omega a_n.
\end{align}

To quantify this effect, we adopt a more convenient, yet equivalent, time-domain description. We consider the readout of an exponentially decaying wavepacket.Suppose the coupler transmittance is fixed at $T = T_0$ from $t = 0$. The output wavepacket is then
\begin{align}
    g_{\mathrm{out}}(t)&=\sqrt{(1-R_0)\tau} \sqrt{R_0}^{\lfloor t/\tau\rfloor\tau}
\end{align}
where $R_0=1-T_0$ and $\tau = L/c$ is the round-trip time. In the single-mode limit, this is approximated by
\begin{align}
g_{\mathrm{ideal}}(t)&=\sqrt{\frac{-\log{R_0}}{\tau}}\exp(\frac12(\log{R_0})\frac{t}{\tau})
\end{align}
The mode overlap between $g_{\mathrm{out}}$ and $g_{\mathrm{ideal}}$ is
\begin{widetext}
\begin{align}
    \sqrt{M}&=\int_0^\infty g_{\mathrm{out}}(t)g_{\mathrm{ideal}}(t)dt\\
    &=\sqrt{-(1-R_0)\log{R_0}}\sum_{n=0}^\infty \int_{n\tau}^{(n+1)\tau} \sqrt{R_0}^n \exp(\frac12(\log{R_0})\frac{t}{\tau}) dt\\
    &=\sqrt{-(1-R_0)\log{R_0}}\sum_{n=0}^\infty \sqrt{R_0}^n \frac{2}{-\log{R_0}}\qty[\exp(\frac12(\log{R_0})n)-\exp(\frac12(\log{R_0})(n+1))]\\
    &=\sqrt{-(1-R_0)\log{R_0}}\sum_{n=0}^\infty R_0^n \frac{2}{-\log{R_0}}\qty(1-\sqrt{R_0})\\
    &=\sqrt{-(1-R_0)\log{R_0}}\frac{1}{1+\sqrt{R_0}} \frac{2}{-\log{R_0}}\\
    &=\sqrt{\frac{T_0}{-\log(1-T_0)}}\frac{2}{1+\sqrt{1-T_0}}\\
    &=1-\frac{T_0^2}{96}+\mathcal{O}(T_0^3)
\end{align}
\end{widetext}

This mode overlap $M$ represents the write/readout efficiency accounting for multimode effects. Figure~\ref{fig:memory_sim}\textbf{d} plots $M$ as a function of $T_0$. The mode overlap remains high for small $T_0$; in our experiment $T_0 = 0.3$, corresponding to a mode overlap of \SI{99.7}{\%}.

\end{document}